\documentclass[prl,twocolumn,showpacs,superscriptaddress,preprintnumbers]{revtex4}
\usepackage{graphicx}
\usepackage{dcolumn}
\usepackage{bm}

\begin{document}

\preprint{}
\title{Individual Mechanisms of Nuclear Spin Decoherence in a Nanoscale GaAs
NMR Device}
\author{Go Yusa}
\email{yusa@NTTBRL.jp}
\homepage{http://www.brl.ntt.co.jp/people/yusa/}
\affiliation{NTT\,Basic\,Research\,Laboratories,\,NTT\,Corporation,\,3-1%
\,Morinosato-Wakamiya,\,Atsugi\,243-0198,\,Japan}
\affiliation{SORST-JST, 4-1-8 Honmachi, Kawaguchi, Saitama 331-0012, Japan}
\author{Norio Kumada}
\email{kumada@will.brl.ntt.co.jp}
\affiliation{NTT\,Basic\,Research\,Laboratories,\,NTT\,Corporation,\,3-1%
\,Morinosato-Wakamiya,\,Atsugi\,243-0198,\,Japan}
\author{Koji Muraki}
\affiliation{NTT\,Basic\,Research\,Laboratories,\,NTT\,Corporation,\,3-1%
\,Morinosato-Wakamiya,\,Atsugi\,243-0198,\,Japan}
\author{Yoshiro Hirayama}
\affiliation{NTT\,Basic\,Research\,Laboratories,\,NTT\,Corporation,\,3-1%
\,Morinosato-Wakamiya,\,Atsugi\,243-0198,\,Japan}
\affiliation{SORST-JST, 4-1-8 Honmachi, Kawaguchi, Saitama 331-0012, Japan}
\date{Version: \today }
\date{\today}

\begin{abstract}
We study decoherence of nuclear spins in a nanoscale GaAs device based on
resistively detected nuclear magnetic resonance (NMR). We demonstrate how
the spin echo technique can be modified for our system, and this is compared
to the damping of Rabi-type coherent oscillations. By selectively decoupling
nuclear-nuclear and electron-nuclear spin, we determine decoherence rates
due to individual mechanisms, namely, direct or indirect dipole coupling
between different or like nuclides and electron-nuclear spin coupling. The
data reveal that the \textit{indirect} dipole coupling between Ga and As
mediated by conduction electrons has the strongest contribution, whereas the 
\textit{direct} dipole coupling between them has the smallest, reflecting
the magic angle condition between the As-Ga bonds and the applied magnetic
field.
\end{abstract}

\maketitle




Nuclear magnetic resonance (NMR) has attracted considerable renewed interest
owing to its suitability for applications in quantum computation and
investigating its pertinent physics \cite{Nielsen}. At present, the most
powerful quantum computer demonstrated with the largest number of qubits is
based on molecules suspended in liquid solution \cite{Vandersypen}. However,
liquid-state NMR lacks scalability, and solid-state devices, in which
preferably microscopic quantities of nuclear spins are manipulated, are
required for implementing realistic scalable quantum computers. In this
context, all electrical control of nuclear spins has recently been
demonstrated in a nanoscale GaAs NMR device\cite{YusaNature}, in which
nanoscopic quantities of nuclear spins are detected resistively by taking
advantage of coupling between nuclear spins and conduction electrons.

In solids, however, decoherence time $T_{2}$, the most important measure of
performance of a qubit, is usually very short compared with $T_{2}$ of
molecules in liquid due to local fluctuations in the strong direct dipole
coupling drastically degrading $T_{2}$. A powerful technique commonly used
to improve $T_{2}$ in solid-state NMR is magic-angle sample spinning \cite%
{Slichter}. However, the technique is not suitable for integrated systems as
spinning imposes crippling geometrical restrictions, and alternative
measures must be sought.

On top of direct dipole coupling, another source of decoherence is through
electron-nuclear spin coupling. The harnessing of this, is of particularly
far-reaching importance, as such coupling is not only a key candidate for
connecting nuclear spin device elements but also a source of decoherence for
electron spin based qubit systems \cite{Khaetskii,Tayler}.

\begin{figure}[t]
\includegraphics[width=0.8\linewidth,clip]{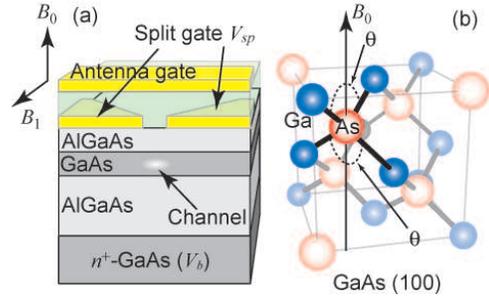}
\caption{(color online) (a) Schematic illustration of the device structure.
The gap between the split gate electrodes is $300$ nm (b) A unit cell of
GaAs. The [100] direction is parallel to the static magnetic field $\mathbf{B%
}_{0}$. An As atom and its nearest neighbor Ga atoms are highlighted.}
\label{fig1}
\end{figure}

In this Letter, we examine decoherence mechanisms of nuclear spins in a
nanoscale GaAs device. We first demonstrate how the spin-echo technique can
be modified for use in the nanoscale device in order to extract the
decoherence time free from inhomogeneous contributions, and compare the
results to the damping of Rabi-type oscillations and discuss their
relationship. Then, by selectively employing heteronuclear decoupling in
combination with a new electron-nuclear spin decoupling technique our novel
device allows, we determine decoherence rates due to individual mechanisms,
namely \textit{direct} or \textit{indirect} dipole coupling between hetero-
or homonuclei and electron-nuclear spin coupling. The data reveal that 
\textit{indirect} dipole coupling between Ga and As mediated by conduction
electrons is the strongest, whereas \textit{direct} dipole coupling between
them makes the smallest contribution, as an As atom and its nearest neighbor
Ga atoms satisfy the magic angle condition \cite{Slichter}.

In order to control and detect nuclear spin states by electron-nuclear spin
coupling \cite{Wald,Kronmuller98,SmetPRL,Gammon,Hashimoto,Machida,Salis,Ono}%
, we use the fractional quantum Hall regime around Landau-level filling
factor $\nu =2/3$, in which coupling of nuclear spins to conduction
electrons through contact hyperfine interactions is known to be pronounced 
\cite{Kronmuller98,SmetPRL,Hashimoto}. In the present study, we use a pair
of split gates to constrict electrons in a channel formed in a $20$-nm
AlGaAs/GaAs/AlGaAs quantum well [Fig. 1(a)]. Then, by applying a current $I_{%
\text{sd}}$ above a certain threshold, nuclear spins can be polarized
selectively in the constricted region \cite{YusaPRB,YusaNature}. When a
pulsed radio-frequency (r.f.) field is applied from the antenna gate [Fig.
1(a)], the magnetic component $B_{1}$ of the r.f. field manipulates nuclear
spins coherently, producing a change in longitudinal magnetization $M_{z}$
of nuclear spins, which we detect through the four-termial resistance $R$ of
the device \cite{detection}. As we have shown previously, $\mathit{\Delta }R$%
, the change in $R$ before and after the r.f. pulse, is proportional to $%
\Delta M_{z}$, the change in $M_{z}$ \cite{YusaNature}. Here, we define $%
\mathit{\Delta }R$ as positive when $R$ drops after the r.f. pulse. All the
measurements were performed with $B_{1}$ of $0.3$ mT at a temperature of $%
\sim 0.1$ K and at static magnetic field $B_{0}=6.3$ T with the split gate
voltage $V_{\text{sp}}$ and $I_{\text{sd}}$ of $-0.25$ V and $9$ nA,
respectively, unless otherwise specified.

\begin{figure}[t]
\includegraphics[width=0.8\linewidth,clip]{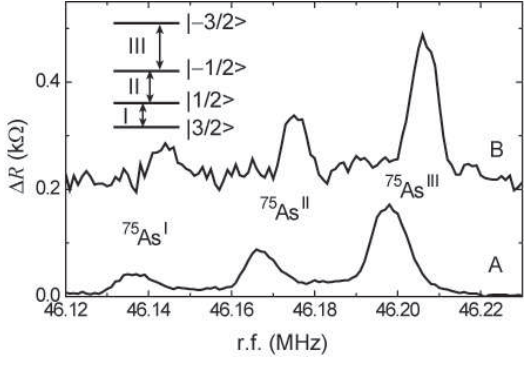}
\caption{NMR spectra A: without and B: with electron-nuclear spin
decoupling. The Larmor frequencies for A and B are $f_{0}^{e}=$ $46.1665$
and $f_{0}=$ $46.1752$ MHz, respectively, giving a Knight shift $\protect%
\delta =f_{0}-f_{0}^{e}$ of $8.7$ kHz. The quadrupolar frequency, $\Delta /h$%
, is $15.33$ kHz. The FWHM of $^{75}$As$^{\text{III}}$ in A and B are $10.9$
and $8.3$ kHz, respectively. B has been offset ($=200$ $\Omega $) for
clarity. Heteronuclear decoupling is performed for these measurements.}
\label{fig2}
\end{figure}

The lower trace in Fig. 2 (A) shows $\Delta R$ measured as the r.f.
frequency was scanned \cite{decouple}. The pulse width $\tau _{\text{P}}$
was set to $0.13$ ms, which corresponds to a $\pi $-pulse of the transition
between states $|-1/2>$ and $|-3/2>$, denoted as $^{75}$As$^{\text{III}}$
(inset to Fig. 2). These three peaks correspond to transitions between
adjacent spin levels in the spin-3/2 system of $^{75}$As, which are
spectrally split by an additional quadrupole interaction. In the following,
we focus on the transition $^{75}$As$^{\text{III}}$ and examine its nuclear
spin dynamics.

We first demonstrate how the spin echo technique,\ a conventional method to
determine $T_{2}$ in standard NMR \cite{Slichter}, can be modified to be
compatible with our $M_{z}$-detection method and examine the $T_{2}$ time.
In spin echo for standard NMR, a sequence of pulses is applied, which
consists of an exciting $\pi /2$-pulse and a refocusing $\pi $-pulse
separated by time interval $\tau $ (denoted by $\pi /2$-$\tau $-$\pi $),
where the refocusing pulse is used to cancel out inhomogeneous contributions
to the decoherence, allowing the intrinsic decoherence to be extracted \cite%
{Slichter}. In contrast to standard NMR, which detects the echo signal as
free induction decay of \textit{transverse} magnetization $M_{xy}$ \cite%
{Slichter}, our scheme probes \textit{longitudinal} magnetization $M_{z}$ 
\cite{YusaNature}, which then requires an additional detection pulse (either 
$\pi /2$ or $3\pi /2$) to convert $M_{xy}$ to $M_{z}$ at the end of the
pulse sequence. Figure 3(a) shows $\mathit{\Delta }R$ obtained by two
different pulse sequences (A: $\pi /2$-$\tau $-$\pi $-$\tau $-$\pi /2$ and
B: $\pi /2$-$\tau $-$\pi $-$\tau $-$3\pi /2$) as a function of $\tau
^{\prime }$, the total time length of the pulse sequence [inset to Fig.
3(a)]. The $\tau ^{\prime }$ dependence of $\mathit{\Delta }R$ can be well
described by exponential decay functions $\propto 1\pm \exp (-\tau ^{\prime
}/T_{2}^{\text{Echo}})$ ($-$ and $+$ for A and B, respectively). The decay
time $T_{2}^{\text{Echo}}$ is estimated to be $\sim 0.75$ ms \cite{decouple}.

\begin{figure}[t]
\includegraphics[width=0.8\linewidth,clip]{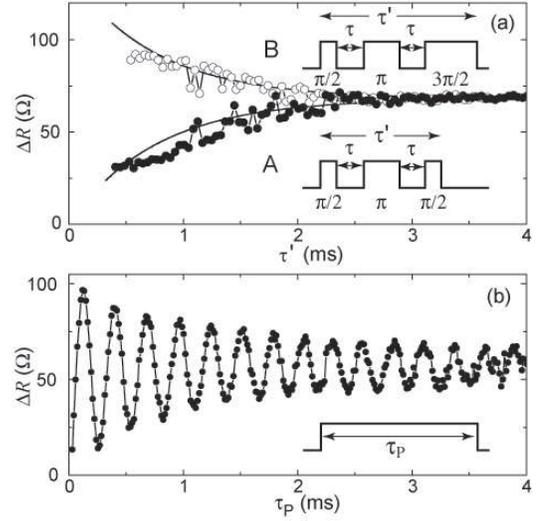}
\caption{(a) Spin echo experiment based on the $M_{z}$-detection method 
\protect\cite{YusaNature} with two different pulse sequences. A: $\protect%
\pi /2$-$\protect\tau $-$\protect\pi $-$\protect\tau $-$\protect\pi /2$ and
B: $\protect\pi /2$-$\protect\tau $-$\protect\pi $-$\protect\tau $-$3\protect%
\pi /2$ (inset), where the time length of the $\protect\pi /2$-pulse, $%
\protect\tau _{\protect\pi /2}$, is $67$ $\protect\mu $s. $\Delta R$ is
plotted as a function of time length $\protect\tau \prime =\protect\tau %
+4 \protect\tau _{\protect\pi /2}$ for A and $\protect\tau \prime =%
\protect\tau +6 \protect\tau _{\protect\pi /2}$ for B. (b) Rabi-type
coherent oscillations of $\Delta R$ by applying single r.f. pulses with
pulse width $\protect\tau _{\text{P}}$ (inset). Heteronuclear decoupling is
performed for this measurement.}
\label{fig3}
\end{figure}

Rabi-type coherent oscillations can be observed when a single pulse with a
varying width $\tau _{\text{P}}$ is applied instead of an echo pulse
sequence. As shown in Fig. 3(b), $\mathit{\Delta }R$ as a function of $\tau
_{\text{P}}$ shows clear oscillations. Such coherent oscillations in Fig.
3(b) clearly persist longer than the echo signal in Fig. 3(a). By fitting an
exponentially damped sine function in Fig. 3(b) we obtain $T_{2}^{\text{Rabi}%
}\sim 1.2$ ms \cite{decouple}, which is longer than $T_{2}^{\text{Echo}}$.
During Rabi-type oscillations, since inhomogeneous contributions to $\mathbf{%
M}$ are spherically symmetric on the Bloch sphere, inhomogeneous
contributions are naturally refocused and are canceled out, and does not
affect $T_{2}^{\text{Rabi}}$, as with $T_{2}^{\text{Echo}}$. However, in
spin echo experiments the magnetization vector $\mathbf{M}$ of nuclear spins
is on the $xy$-plane \cite{Slichter}, whereas in coherent oscillation
measurements $\mathbf{M}$ is always driven by the r.f. field continuously
around the great circle of the Bloch sphere \cite{Nielsen}, and is robust
against decoherence when $\mathbf{M}$ is near the localized states $|-3/2>$
and $|-1/2>$. The value of $T_{2}^{\text{Echo}}$ we obtain is therefore
consistent with expectation and demonstrate that the altered spin-echo
technique successfully measures the decoherence time.

\begin{figure}[t]
\includegraphics[width=0.8\linewidth,clip]{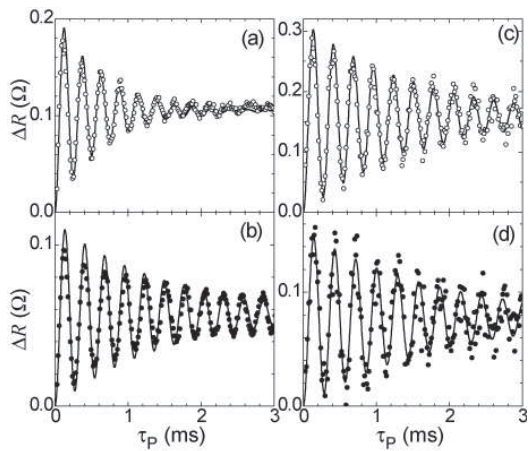}
\caption{Coherent oscillations with and without heteronuclear and
electron-nuclear spin decoupling. The r.f. frequencies used were $46.1985$
MHz for (a) and (b), and $46.207$ MHz for (c) and (d) to allow for the
Knight shift. The broadband CW r.f. radiation used for heteronuclear
decoupling had a magnetic component with peak intensity $\sim 2$ $\protect%
\mu $T and was frequency-modulated by the internal noise of each r.f.
generator. (a) Without any decoupling. (b) With heteronuclear decoupling.
(c)With electron-nuclear spin decoupling. (d) With heteronuclear and
electron-nuclear spin decouplings.}
\label{fig4}
\end{figure}

In the remainder of this paper, we concentrate on evaluating the decoherence
in our nuclear spin qubit. We focus on $T_{2}^{\text{Rabi}}$ as opposed to $%
T_{2}^{\text{Echo}}$ since $T_{2}^{\text{Rabi}}$ ($=T_{2}$, henceforth) is
more directly relevant to quantum computation as it represents the length of
time available for operations \cite{Nielsen}. In order to distinguish
individual decoherence mechanisms, we employ two distinctly different
techniques to decouple $^{75}$As nuclei from their environment.

The first is the following broadband decoupling \cite{Freeman}. In addition
to the r.f. pulse resonant with $^{75}$As$^{\text{III}}$, broadband
continuous-wave (CW) r.f. with center frequencies at Larmor frequencies of $%
^{69}$Ga and $^{71}$Ga ($64.74$ and $82.26$ MHz at $B_{0}=6.3$ T) and
bandwidths of $70$ and $50$ kHz covering the whole quadrupolar-split
spectrum of each Ga isotope are combined together and applied to the point
contact through the antenna gate. Such CW broadband r.f. randomizes $%
^{69,71} $Ga nuclear spins, thereby averaging the inhomogeneous local field
due to heteronuclear dipole coupling. Figure 4(a) and (b) compare coherent
oscillations of $^{75}$As obtained (a) without and (b) with the
heteronuclear decoupling. The data clearly show that the coherent
oscillations persist longer with the decoupling. $T_{2}$ is found to be
improved from $\sim 0.6$ to $1.2$ ms.

The second is the following electron-nuclear spin decoupling. Our device
allows to decouple nuclei from the electron, since electrons in the
point-contact region can be depleted during the manipulation of nuclei with
the r.f. pulse. This was achieved by setting $I_{\text{sd}}$ to zero and
applying a larger negative voltage, $V_{\text{sp}}=-0.8$ V, to the split
gates. We find that the NMR spectrum thus obtained (Line B in Fig. 2)
appears at slightly higher frequencies (by $8.7$ kHz) compared with that
without electron-nuclear spin decoupling (Line A). The shift due to the
presence of conduction electrons, \textit{i.e}., the Knight shift \cite%
{Slichter,Barrett,Kuzma,Khandelwal,Stern}, reflects the effective magnetic
field ($-1.187$ mT \cite{shift}) produced by the spins of the conduction
electrons, which modifies the Zeeman energy of nuclear spins \cite{width}.
Figure 4(c) shows coherent oscillations obtained with this electron-nuclear
spin decoupling, which shows $T_{2}$ to be enhanced to $\sim 1.5$ ms. By
using both of the two decoupling techniques, $T_{2}$ is further extended to $%
\sim 1.8$ ms, as shown in Fig. 4(d).

\begin{table}[t]
\caption{Table I Contributions to the decoherence rate $1/T_{2}$ (ms$^{-1}$)
for each panel of Fig. 4., with measured values.}%
\begin{tabular}{cccccc}
\hline\hline
Fig. 4 & \multicolumn{4}{c}{Contributions} & $1/T_{2}$ \\ \hline
(a) & As-Ga, & As-As, & As-\textit{e}-Ga, & As-\textit{e}-As, As-\textit{e}
& $\sim 1.67$ \\ 
(b) &  & As-As, &  & As-\textit{e}-As, As-\textit{e} & $\sim 0.83$ \\ 
(c) & As-Ga, & As-As, &  &  & $\sim 0.67$ \\ 
(d) &  & As-As, &  &  & $\sim 0.56$ \\ \hline\hline
\end{tabular}%
\end{table}
\begin{table}[t]
\caption{ Individual contributions to the decoherence rate $1/T_{2}$ (ms$%
^{-1} $) \protect\cite{homo}.}%
\begin{tabular}{ccccc}
\hline\hline
& As-Ga & As-\textit{e}-Ga & As-As & As-\textit{e}-As, As-\textit{e} \\ 
\hline
$1/T_{2}$ & $\sim 0.11$ & $\sim 0.73$ & $\sim 0.56$ & $\sim 0.27$ \\ 
\hline\hline
\end{tabular}%
\end{table}

Now we extract decoherence \textit{rates}, $1/T_{2}$, contributed by
individual mechanisms from the set of data in Figs. 4(a)-(d). A comparison
between (a) and (b) reveals an improvement in $1/T_{2}$ by $\sim 0.84$ ms$%
^{-1}$, which can be attributed to direct and indirect heteronuclear dipole
coupling (denoted as As-Ga and As-\textit{e}-Ga). Similarly, by comparing
(a) and (c), the improvement of $1/T_{2}$ by $\sim 1.0$ ms$^{-1}$ can be
ascribed to the mechanisms involving conduction electrons, \textit{i.e}.,
the electron-nuclear spin coupling (denoted as As-\textit{e}) and the 
\textit{indirect} homo- or heteronuclear dipole coupling mediated by
conduction electrons (denoted as As-\textit{e}-As and As-\textit{e}-Ga,
respectively). In (d), where all these mechanisms are eliminated, the
observed $1/T_{2}\sim 0.56$ ms$^{-1}$, can be identified as due to direct
homonuclear dipole coupling (As-As) \cite{others}. Table I shows a summary
of the estimations \cite{homo}. Putting all these results together, we
arrive at decoherence rates due to individual mechanisms as presented in
Table II.

The results obtained above reveal that, compared to other mechanisms, the
heteronuclear direct dipole coupling (As-Ga) only makes a small contribution
to the decoherence, which may seem counterintuitive. In general such direct
dipole coupling, which is inversely proportional to $|\mathbf{r|}^{3}$,
where $\mathbf{r}$ is the vector between two nuclei \cite{Slichter}, is
known to be the strongest \cite{Han}, because Ga atoms are nearest neighbors
to As atoms (with distance $r_{1}=0.433a$. $a=0.565$ nm is the lattice
constant of GaAs \cite{Madelung}.) However, the dipole coupling is also
dependent on the angle $\theta $ between $\mathbf{r}$ and vector $\mathbf{B}%
_{0}$ by a factor $3\cos ^{2}\theta -1$ \cite{Slichter}. When $\mathbf{B}%
_{0} $ points along the [100] direction [see Fig. 1(b)], angles of $sp^{3}$
bonds between As and its nearest neighbor Ga atoms are all $\theta
=54.7347^{^{\circ }}$, which satisfies $3\cos ^{2}\theta -1=$ $0$ (\textit{%
i.e.}, the magic angle). The main contribution to As-Ga, therefore, comes
from Ga nuclei in the third nearest ($r_{3}=1.0897a$) or farther lattice
sites \cite{powder}. In contrast, As-\textit{e}-Ga has the strongest contribution among
other contributions because As and Ga nuclei can couple with the shortest
distance $r_{1}$ via conduction electrons.

In summary, we extracted individual decoherence rates of nuclear spins in a
nanoscale region by introducing electrically controlled decoupling
techniques: heteronuclear and electron-nuclear spin decoupling. We
demonstrated that heteronuclear direct dipole coupling is a less important
factor in determining the decoherence compared with other sources due to the
magic angle of the crystal bonds. This result is applicable to any Zinc
blend and diamond structure when the crystal is placed such that the [100]
direction is parallel to $\mathbf{B}_{0}$. The orientation of crystal will
play an important role for controlling interactions between heteronuclei for
quantum entanglement in multi-qubit quantum computation using heteronuclei.
In our experiment the $T_{2}$ time is extended to $\sim 1.8$ ms by a factor
of $\sim 3$. Further improvements to $T_{2}$ will be allowed by homonuclear
decoupling and more sophisticated heteronuclear decoupling techniques \cite%
{Slichter,Freeman,longT2}.

\begin{acknowledgments}
The authors are grateful to K. Takashina, T. Fujisawa, H. Yamaguchi, K.
Hashimoto, and T. Ota for fruitful discussions. Y. H. is partially supported
by a Grant-in-Aid for Scientific Research from JSPS.
\end{acknowledgments}


\begin{references}

\bibitem{Nielsen} M. A. Nielsen and I. L. Chuang, 
\textit{Quantum Computation and Quantum Information}, 
(Cambridge Univ. Press 2003).

\bibitem{Vandersypen} L. M. K. Vandersypen \textit{et al.}, Nature 
\textbf{414}, 883 (2001).

\bibitem{YusaNature} G. Yusa \textit{et al.},
Nature \textbf{434}, 1001 (2005).

\bibitem{Slichter} C. P. Slichter \textit{%
Principles of Magnetic Resonance} 3rd ed., (Springer, Tokyo, 1989).

\bibitem{Khaetskii} A. V. Khaetskii, D. Loss, and L. Glazman, 
Phys. Rev. Lett. {\bf 88}, 186802 (2002).
 
\bibitem{Tayler}  J. M. Taylor, C. M. Marcus, and M. D. Lukin, 
Phys. Rev. Lett. {\bf 90}, 206803 (2003). 

\bibitem{Kronmuller98} S. Kronm\"{u}ller
\textit{et al.}, Phys. Rev.
Lett. \textbf{81}, 2526 (1998).

\bibitem{SmetPRL} J.
H. Smet \textit{et al.}, Phys. Rev. Lett. \textbf{86}%
, 2412 (2001).

\bibitem{Hashimoto} K. Hashimoto \textit{et al}., Phys. Rev. Lett.
\textbf{88}, 176601 (2002).

\bibitem{Wald} K. Wald \textit{et
al.}, Phys. Rev. Lett. \textbf{73}, 1011 (1994).

\bibitem{Gammon} D. Gammon \textit{et
al.}, Science \textbf{277}, 85 (1997).

\bibitem{Salis}
G. Salis \textit{et al.}, Phys. Rev. 
Lett. \textbf{86}, 2677 (2001).

\bibitem{Machida} T.
Machida \textit{et al.}, Appl. Phys.
Lett. \textbf{82}, 409 (2003).

\bibitem{Ono} 
K. Ono and S. Tarucha, 
Phys. Rev. Lett. \textbf{92}, 256803 (2004).

\bibitem{YusaPRB} G. Yusa \textit{et al.},
Phys. Rev. B \textbf{69}, 161302(R) (2004).

\bibitem{detection} For detection of nuclear spin polarization, we 
take advantage of competition between spin-polarized and unpolarized 
ground states at this $\nu$, which makes $R$ 
highly sensitive to small changes in the electronic
 Zeeman energy $E_{z}$ and hence to the nuclear polarization 
that modifies $E_{z}$ through the hyperfine field it produces. 

\bibitem{decouple}
Heteronuclear decoupling (see text) is performed for Fig. 2 and Fig. 3.

\bibitem{Freeman} R. Freeman \textit{%
Spin Choreography: Basic
Steps in High Resolution Nmr.} (Oxford Univ. Press Oxford, 1998). 

\bibitem{Barrett} S. E. Barrett \textit{et al.}, Phys. Rev. Lett. 
\textbf{74}, 5112 (1995).

\bibitem{Kuzma} N. N. Kuzma \textit{et al.},
Science \textbf{281}, 686 (1998).

\bibitem{Khandelwal} P. Khandelwal 
\textit{et al.}, Phys. Rev. Lett.
\textbf{81}, 673 (1998).

\bibitem{Stern} O. Stern \textit{et al.}, Phys. Rev. B 
\textbf{70}, 075318 (2004).


\bibitem{shift} It is estimated by the ratio between the Knight shift 
and the gyromagnetic ratio $\gamma=7.3294 \times 10^6$ T$^{-1}$s$^{-1}$ of $^{75}$As, 
which is taken from Fig. 2.

\bibitem{width}
In our system $T_2$ is sufficiently long so that 
the peak width is not limited by $1/T_2$, but by off-resonance features, 
which also depend on $B_1$.
Simulations of $\Delta R$ 
neglecting decoherence ($1/T_2=0$) also show finite width, comparable to 
the measured width (See Fig. 4 in Ref. \cite{YusaNature}). 
Therefore, only a slight difference in widths is seen in Fig. 2 between the 
Knight-shifted and the original peaks, in stark contrast to NMR spectra in conventioanl 
two-dimensional electron systems \cite{Barrett,Kuzma,Khandelwal}.

\bibitem{others} Other contributions to decoherence, such as 
fluctuations in quadrupolar interactions, are included here.

\bibitem{homo} Since we do not perform homonuclear decoupling of $^{75}$As, 
As-\textit{e}-As and As-\textit{e} are not separable.

\bibitem{Han} See for example for GaAs: 
O. H. Han, H. K. C. Timken, and E. Oldfield, 
J. Chem. Phys. \textbf{89}, 6046 (1988).  

\bibitem{Madelung} O. Madelung \textit{ed}. 
\textit{Seimiconductors - Basic Data} 3nd ed., 
(Springer-Verlag Berlin, 1996).

\bibitem{powder} For standard solid-state NMR, samples are in powder form or
bundles of crystals.
In such samples the crystal orientation is randomized
and dipole coupling dominates 
NMR peak widths and therefore decoherence of nuclear
spins as in \cite{Han}.

\bibitem{longT2}
Recently an extremely long $T_2$ time ($\sim25$ s) has been reported in isotopically 
diluted spin-1/2 $^{29}$Si. This is obtained in a spin-0 bulk Si crystal 
using high-power decoupling and magic-angle sample spinning. 
See T. D. Ladd \textit{et al.}, Phys. Rev. B \textbf{71}, 014401 (2005).

\end{references}

%

\end{document}